\documentstyle[12pt]{article}\textheight
230mm\textwidth 165mm 
\newcommand{\bi}{\bibitem}
\newcommand{\be}{\begin{eqnarray}}
\newcommand{\ee}{\end{eqnarray}}
\newcommand{\nn}{\nonumber}
\catcode`\@=11
\def\lsim{\mathrel{\mathpalette\@versim<}}
\def\gsim{\mathrel{\mathpalette\@versim>}}
\def\@versim#1#2{\vcenter{\offinterlineskip
\ialign{$\m@th#1\hfil##\hfil$\crcr#2\crcr\sim\crcr } }}
\catcode`\@=12

\begin{document}
\pagestyle{empty}

\noindent
\hspace*{10.7cm} \vspace{-3mm}  HIP-98-06/TH\\
\hspace*{10.7cm} \vspace{-3mm} KANAZAWA-98-02\\

\vspace{0.3cm}
\noindent
\hspace*{10.7cm} February 1998

\vspace{0.3cm}

\begin{center}
{\Large\bf   FURTHER ALL-LOOP RESULTS  \vspace{-1mm}
 \\ IN SOFTLY-BROKEN SUPERSYMMETRIC GAUGE THEORIES}
\end{center} 

\vspace{1cm}

\begin{center}
{\sc Tatsuo Kobayashi}$\ ^{(1)}$, 
{\sc Jisuke Kubo}$\ ^{(2),*}$ and
{\sc George Zoupanos}$\ ^{(3),**}$  
\end{center}
\begin{center}
{\em $\ ^{(1)}$ 
Department of Physics,  High Energy Physics Division, 
University of Helsinki \vspace{-2mm}\\ and \vspace{-2mm}\\
Helsinki Institute of Physics,
FIN-00014 Helsinki, Finland} \\
{\em $\ ^{(2)}$ 
Department of Physics, 
Kanazawa \vspace{-2mm} University,
Kanazawa 920-1192, Japan } \\
{\em $\ ^{(3)}$ 
Institut f\" ur Physik, Humboldt-Universit\" at zu Berlin, \vspace{-2mm}
D-10115 Berlin, Germany}
\end{center}

\vspace{0.5cm}
\begin{center}
{\sc\large Abstract}
\end{center}

\noindent
It is proven that 
the recently found, 
renormalization-group invariant \vspace{-2mm} sum rule for the soft
scalar masses in softly-broken
$N=1$ supersymmetric gauge-Yukawa unified theories \vspace{-2mm}
can be extended to all orders in perturbation theory. \vspace{-2mm}
In the case of finite unified theories, 
the sum rule ensures \vspace{-2mm} 
the all-loop finiteness in the soft supersymmetry breaking sector.
As a byproduct the exact $\beta$ function for 
the soft scalar masses in the \vspace{-2mm}
Novikov-Shifman-Vainstein-Zakharov (NSVZ) 
scheme for softly-broken supersymmetric
QCD is obtained.
It is also found that \vspace{-2mm}
the singularity appearing in the sum rule 
in the NSVZ scheme exactly coincides \vspace{-2mm}
with that which 
has been  previously found
in a certain class of 
superstring models in \vspace{-2mm}
which the massive string states are organized 
into $N = 4$ supermultiplets.

\vspace*{1cm}
\footnoterule
\vspace*{2mm}
\noindent
$^{*}$Partially supported  by the Grants-in-Aid
for \vspace{-3mm} Scientific Research  from the Ministry of
Education, Science 
and Culture \vspace{-3mm}  (No. 40211213).\\
\noindent
$ ^{**}$
On leave from:
Physics Dept., Nat. Technical University, GR-157 80 
\vspace{-3mm} Zografou,\\
Athens, Greece.
Partially supported  by the E.C. projects, \vspace{-3mm}
FMBI-CT96-1212 and  ERBFMRXCT960090,
the Greek projects, PENED95/1170; 1981.

\newpage
\pagestyle{plain}
\section{Introduction}

The plethora of free parameters of the, very successful otherwise, 
Standard Model (SM), can be interpreted as signaling the existence of a
more fundamental Physics picture in higher scales,
whose remnants appear as
free parameters in the SM. In fact after several decades of experience in
searching for such a fundamental theory, which in principle could explain
everything that is observed today in terms of very few parameters, it
seems more realistic to expect that only parts of the fundamental theory
are uncovered at various higher scales; maybe the full fundamental theory
can only be found close to the Planck scale. 
The usual theoretical strategy
to search for new Physics beyond the SM is to construct more symmetric
theories, e.g. 
Grand Unified Theories (GUTs) at higher scales and subsequently test their
predictions against the measured low energy parameters. A representative
candidate for carrying some of the information of the fundamental theory
at intermediate scales is the $N=1$ globally 
supersymmetric $SU(5)$ GUT,
given its predictive power for certain 
low energy free parameters of the
SM.

 In our recent studies \cite{finite1}--\cite{ksz1} 
\footnote{For an extended discussion
 and a complete list of references, see ref. \cite{kmz3}.},
we have developed
another complementary strategy in searching for a more fundamental theory
possibly at Planck scale and its consequences that could be missing in
ordinary GUTs. Our method consists of hunting for renormalization group
invariant (RGI) relations among couplings  holding below the Planck scale
and which  therefore are exactly preserved down to the GUT scale. This
programme applied in the dimensionless couplings of supersymmetric GUTs
such as gauge and Yukawa couplings had already certain success by
predicting correctly, among others, the top quark mass in the 
finite \cite{finite1,kmoz1}
and in the minimal  \cite{kmz1,kmoz1} $N = 1$ supersymmetric
$SU(5)$-GUTs.

 An impressive aspect of the RGI relations is that one can guarantee their
validity to all-orders in perturbation theory by studying the uniqueness
of the resulting relations at one-loop, as was proven in the early days
of the programme of {\em reduction of couplings} \cite{zim1}.

 Although supersymmetry seems to be an essential feature for a successful
realization of the above programme, its breaking has to be understood too
in this framework,  which has the ambition to supply the SM
with predictions for several of its free
parameters. Therefore, the search for RGI 
relations was naturally extended
to the soft supersymmetry breaking (SSB) sector 
of these theories \cite{jack2,kmz2}, which
involve parameters with dimension one and two. 
In the case of nonfinite theories, the method to prove
the existence of reduction of couplings to all-loop
\cite{zim1}--\cite{ksz1}
can be easily extended  for
the RGI relations among dimensional parameters \cite{kmz2}  if 
use of a mass-independent renormalization scheme (RS)
is assumed \footnote{The proof is also possible 
without any assumption on a particular RS \cite{zim2}.}.
In contrast  to this, for the case of finite theories 
the elegant way of ref. \cite{lucchesi1} to show finiteness 
(which is based on a consideration of renormalization of 
certain anomalies)  cannot be simply applied;
reduction of couplings is merely one of the  conditions
for finiteness.
The proof of the all-order finiteness 
is certainly less involved to be performed in a particular RS
in which various properties of the RG functions 
are known and can be assumed \cite{ermushev1}. Using the recent
results  \cite{yamada1}--\cite{avdeev1} on the renormalization
properties of the SSB sector in the 
supersymmetric version of the minimal subtraction scheme,
Kazakov \cite{kazakov1} has pursued that line of the thought
and shown the finiteness in the SSB 
sector \footnote{Finiteness in this sector in
lower orders are shown in refs. \cite{jones1,jack1}}.
Soon later Jack, Jones and Pickering \cite{jack4}
have generalized Kazakov's idea \cite{kazakov1} so as to find RGI
relations among the SSB parameters in the nonfinite case.

Note that in the formulation of 
 references above
the SSB parameters are expressed in terms
of the unified gauge coupling $g$ and the
unified gaugino mass parameter $M$ only,
which may appear as a  too strong constraint on the SSB sector
for a given phenomenological model.
Therefore, there has been attempts \cite{kkk1,kkmz1} 
to relax this constraint
without loosing RGI.
An interesting 
observation resulting from the 
independent analysis of the SSB sector of a $N=1$
supersymmetric gauge-Yukawa unified theory is the 
existence of a RGI sum rule
for the soft scalar- masses
in lower orders; in  one-loop for the nonfinite case 
\cite{kkk1} and in 
two-loop for the finite case \cite{kkmz1}. The sum rule appears to have
significant phenomenological consequences and in particular manages to
overcome the unpleasant predictions of the previously known ``universal''
finiteness condition for the soft scalar masses 
\cite{jones1,jack1}. The
universal soft scalar masses apart from their simplicity they were
appealing for a number of reasons (a) they are part of the constraints
that preserve finiteness up to two-loop 
\cite{jones1,jack1}, (b) they appear to be
RGI under a certain constraint, known as the $P =1/3 Q$ 
condition \cite{jack2}, in more
general supersymmetric gauge theories, and (3) they appear in the dilaton
dominated supersymmetry breaking superstring scenarios 
\cite{ibanez1}. In the latter case,
since the dilaton couples in a universal manner to all particles the
universality of soft scalar masses appears as a quite model independent
feature. Unfortunately, further studies have exhibited a number of problems
attributed to the universality of soft scalar masses. For instance
(1) in finite unified theories the universality leads to a charged
particle, the superpartner of $\tau$, the s-$\tau$, to be the lightest 
supersymmetric particle \cite{yoshioka1,kkmz1}, 
(2) the MSSM with universal soft scalar masses is inconsistent with
radiative electroweak symmetry breaking \cite{brignole1}
 and (3) worst
of all the dilaton dominated limit leads to charge and/or colour breaking
minima deeper than the standard vacuum \cite{casas1}. 
Therefore, the sum
rule is a  welcome possibility. 
Furthermore, it was shown that
the same sum rule is satisfied in a certain class 
of 4D orbiford models,
at least at the tree-level for the nonfinite \cite{kkk1}
and in two-loop order for the finite case \cite{kkmz1} 
if the massive string states are organized into $N =4$
supermultiplets so that they do not 
contribute to the quantum modification
of the gauge kinetic function \cite{kubo1}.

The purpose of
the present paper is to prove the existence of the 
RGI soft scalar-mass sum
rule to all-orders for the nonfinite as well as for the finite case, based
on the recent developments on the renormalization properties of 
the SSB sector of the $N=1$ supersymmetric
 gauge theories. As an interesting
byproduct we obtain the exact $\beta$ function for 
the soft scalar masses in the 
Novikov-Shifman-Vainstein-Zakharov (NSVZ)
scheme \cite{novikov1} for softly-broken $N=1$ supersymmetric
QCD.

\section{Recent results on the renormalization of the SSB parameters}

Most of the recent interesting progress 
\cite{hisano1}--\cite{kazakov1}, \cite{jack4} on the
renormalization properties of the SSB parameters 
is based conceptually and
technically on the work of ref. \cite{yamada1}. 
In ref. \cite{yamada1} the 
powerful supergraph method \cite{delbourgo1} for studying supersymmetric
theories has been applied to the softly-broken ones
by using the ``spurion'' external space - time 
independent superfields \cite{girardello1}.
In the latter method a softly-broken supersymmetric gauge theory is
considered as a supersymmetric one 
in which the various parameters such as
couplings and masses have been promoted to external superfields that
acquire ''vacuum expectation values''. 
Based on this method the relations
among the soft term renormalization and that of an unbroken
supersymmetric gauge theory have been derived.

 To be more specific, following the notation of
 ref. \cite{jack4}, in an $N=1$
supersymmetric gauge theory with superpontential
\be
W(\Phi) &= &\frac{1}{6} Y^{ijk} \Phi_i \Phi_j \Phi_k + 
\frac{1}{2} \mu^{ij} \Phi_i
\Phi_j 
\ee
the SSB part $L_{SSB}$ can be written as \cite{yamada1}:
\be
L(\Phi,W) &=& - \left( ~\int d^2\theta\eta (  \frac{1}{6} 
 h^{ijk} \Phi_i \Phi_j \Phi_k +  \frac{1}{2}  b^{ij} \Phi_i \Phi_j 
+  \frac{1}{2}  MW_A^\alpha W_{A\alpha} )+
\mbox{h.c.}~\right)\nn\\
& &-\int d^4\theta\tilde{\eta} \eta \overline{\Phi^j}                   
(m^2)^i_j(e^{2gV})_i^k \Phi_k~,
\ee
where $\eta = \theta^2$, 
$\tilde{\eta} = \tilde{\theta}^2$ are the external
spurion superfields and $\theta$, $\tilde{\theta}$ 
are the usual grasmannian
parameters, and $M$ is the gaugino mass.
The $\beta$ functions of the $M, h$ and $m^2$
parameters  are found to be:
\be
\beta_M &=& 2{\cal O}\left({\beta_g\over g}\right)~,
\label{betaM}\\
\beta_h^{ijk}&=&\gamma^i{}_lh^{ljk}+\gamma^j{}_lh^{ilk}
+\gamma^k{}_lh^{ijl}-2\gamma_1^i{}_lY^{ljk}
-2\gamma_1^j{}_lY^{ilk}-2\gamma_1^k{}_lY^{ijl}~,\\
(\beta_{m^2})^i{}_j &=&\left[ \Delta 
+ X \frac{\partial}{\partial g}\right]\gamma^i{}_j~,
\label{betam2}\\
{\cal O} &=&\left(Mg^2{\partial\over{\partial g^2}}
-h^{lmn}{\partial
\over{\partial Y^{lmn}}}\right)~,
\label{diffo}\\
\Delta &=& 2{\cal O}{\cal O}^* +2|M|^2 g^2{\partial
\over{\partial g^2}} +\tilde{Y}_{lmn}
{\partial\over{\partial
Y_{lmn}}} +\tilde{Y}^{lmn}{\partial\over{\partial Y^{lmn}}}~,
\label{delta}\ee
where $(\gamma_1)^i{}_j={\cal O}\gamma^i{}_j$, 
$Y_{lmn} = (Y^{lmn})^*$, and 
\be
\tilde{Y}^{ijk}&=&
(m^2)^i{}_lY^{ljk}+(m^2)^j{}_lY^{ilk}+(m^2)^k{}_lY^{ijl}~.
\ee

Note that the $X$ term in (\ref{betam2}) is explicitly known only
in the lowest order \cite{jack1,jack5}:
\be
X^{(2)} &=& -\frac{S g^3}{8\pi^2}~,~
S\delta_{AB}=(m^2)^k{}_l(R_AR_B)^l{}_k
-|M|^2 C(G)\delta_{AB}~.
\label{x2}
\ee
We do not consider the $b$ parameters
in the following discussions, because they do not enter into the
$\beta$ functions of the other quantities at all. Moreover they are
finite if the other quantities are finite.

In order to express the $h$ and $m^2$ parameters in terms of $g$ and $M$
in a RG invariant way, we have to solve the set of coupled reduction
equations \cite{zim1,osz1,ksz1}. 
The key point in the strategy of 
refs. \cite{kazakov1,jack4} to solve the reduction
equations is the assumption that the differential operators
${\cal O}$ and $\Delta$ given in eqs. (\ref{diffo})
and  (\ref{delta})
become total derivative operators
on the RG invariant surface 
which is defined by the solution
of the reduction solutions.
Although we consider this assumption as a subtle one and the extent of its
validity requiring further clarification, we accept it throughout our
analysis.

Observe that the $\beta$ functions of the SSB parameters
are obtained 
by applying the differential operators,
${\cal O}$ and $\Delta$, on 
the RG functions, $\beta_g$ and $\gamma^j{}_i$, of the unbroken
theory, and
note next that in a finite theory 
 $Y^{ijk}$ is a power series of $g$  and that
 $\beta_g$ as well as $\gamma^j{}_i$ have
to identically vanish.
But in general we expect that
\be
\frac{\partial \gamma^j{}_i(g, Y, Y^*)}{\partial Y}
~\left|_{Y=Y(g),Y^*=Y^*(g)}\right. \neq 0~~\mbox{or}~~
\frac{\partial \gamma^j{}_i(g, Y, Y^*)}{\partial g}
~\left|_{Y=Y(g),Y^*=Y^*(g)}\right. \neq 0~,
\ee
even if $\gamma^j{}_i(g, Y(g), Y^*(g))$ vanishes.
However, one easily sees that
\be
& &\frac{d \gamma^j{}_i}{d g}(g, Y=Y(g), Y^*=Y^*(g))\nn\\
& & =
\frac{\partial \gamma^j{}_i(g, Y, Y^*)}{\partial g}
~\left|_{Y=Y(g),Y^*=Y^*(g)}\right.
 +\frac{\partial \gamma^j{}_i(g, Y, Y^*)}{\partial Y}
~\left|_{Y=Y(g),Y^*=Y^*(g)}\right. \frac{d Y(g)}{d g}\nn\\
& &+\frac{\partial \gamma^j{}_i(g, Y, Y^*)}{\partial Y^*}
~\left|_{Y=Y(g),Y^*=Y^*(g)}\right. \frac{d Y^*(g)}{d g} =0~,
\ee
if $\gamma^j{}_i(g, Y=Y(g), Y^*=Y^*(g))=0$.
Kazakov \cite{kazakov1} examining the finite case was searching for
a RG invariant surface
on which the differential operators 
${\cal O}$ and $\Delta$ 
can be written as total derivative terms.

In ref. \cite{jack4} the general case has been considered and has been
further  assumed that
\be
\gamma^j{}_i &=& \gamma_i \delta^j{}_i~,
\label{as1}\\
(m^2)^j{}_i &=& m^2_i \delta^j{}_i~,
\label{as2}\\
Y^{ijk}\frac{\partial}{\partial Y^{ijk}} 
&=& Y^{*ijk}\frac{\partial}{\partial Y^{*ijk}}~
\mbox{on the space of the RG functions}~,
\label{as3}
\ee
and has been shown that  if
\be
h^{ijk} &=& -M (Y^{ijk})'
\equiv -M \frac{d Y^{ijk}(g)}{d \ln g}~,
\label{h}\\
  m^2_i &=&
|M|^2 \{~(1+\tilde{X}(g)) (g/\beta_g) (\gamma_i (g))'+
\frac{1}{2}[(g/\beta_g)\gamma_i (g)]'~\}
\label{m2}
\ee
are satisfied, then the differential operators 
${\cal O}$ and $\Delta$ can be written as
\be
{\cal O} &=& \frac{M}{2}
\frac{d }{d \ln g}~,
\label{diffo2}\\
\Delta &=& |M|^2[~
\frac{1}{2}\frac{d^2}{d (\ln g)^2}
+ (1+\tilde{X}(g)/g)\frac{d}{d \ln g}~]~,
\label{delta2}
\ee
where
\be
g\tilde{X}(g) &=&\frac{1}{|M|^2} X(g, Y(g), Y^*(g),
h(M,g),h^*(M,g),m^2(|M|^2,g))~.
\label{xtilde}
\ee
 Eqs. 
(\ref{diffo2}) and (\ref{delta2}) can be derived from
\be
\frac{d \ln Y^{ijk}}{d \ln g}=(\ln Y^{ijk})'=
(g/\beta_g)[\gamma_i (g)+\gamma_j(g)+\gamma_k (g)]~,
\label{lny}
\ee
which follows assuming the reduction equation
\be
\beta_g \frac{d Y^{ijk}(g)}{d g}=
\beta^{ijk}=Y^{ijk}(g)[\gamma_i (g)+\gamma_j(g)+\gamma_k (g)]~.
\ee
Note that so far eq. (\ref{h}) is  a solution
of the reduction equation
(i.e. RG invariant), but  eq. (\ref{m2}) is not.
At the final step, Jack {\em et al.} in ref. \cite{jack4} require
that eq. (\ref{m2}), too, is RG invariant, which fixes 
$\tilde{X}(g)$ uniquely 
up to a term related to an integration
constant. This integration constant term is then  fixed 
by comparing it with the lowest order result in eq. (\ref{x2}).
They found
\be
\tilde{X}(g) &=&\frac{1}{2}(\ln (\beta_g /g))'-1~.
\label{xg}
\ee
Note that there is no perturbative
computation of $X$ beyond two-loop.
Therefore eq. (\ref{xg}) may be understood as a prediction of
perturbative computation of $X$.
If one inserts $\tilde{X}$ above into
eq. (\ref{m2}), one obtains
\be
m^2_i &=&
\frac{1}{2}|M|^2 (g/\beta_g) (\gamma_i (g))'
\label{m21}
\ee
which together with (\ref{h}) is
the final result of ref. \cite{jack4}.

\section{New results}
Next let us consider the sum rules for soft scalar masses 
\cite{kkk1,kkmz1}.
In turn, we  assume neither  (\ref{m2})
nor   (\ref{m21}).
But we assume that $Y^{ijk}$ and $h^{ijk}$ are
already reduced, where $h^{ijk}$ is given 
in eq. (\ref{h}), as well as that (\ref{as1}) --(\ref{as3}) hold.
Suppose  that the sum rule
takes the form
\be
m_i^2+m_j^2+m_k^2 &=&
|M|^2 F^M_{ijk}(g)+ \sum_\l m_\l^2F^\l_{ijk} (g)~.
\label{sum1}
\ee
We require, as in ref. \cite{kazakov1,jack4}, that
$\Delta$ acting on $\gamma_i$
can be written as  a total derivative operator, and we
find that
\be
F^M_{ijk}(g)&=&(1+\tilde{X}^M(g))(\ln Y^{ijk})'
+\frac{1}{2}(\ln Y^{ijk})''~,~
F^\l_{ijk} (g)=
\tilde{X}^\l(g)(\ln Y^{ijk})'~
\label{F}
\ee
have to be satisfied, where
\be
|M|^2 g \tilde{X}^M(g)+
\sum_\l m_\l^2 g \tilde{X}^\l (g) &=&
X(g, Y(g), Y^*(g),
h(M,g),h^*(M,g),m^2) ~.
\label{xmxl}
\ee
Then we have
\be
\beta_{m^2_i} &=&\Delta \gamma_i \nn\\
& =&
 \{|M|^2[~
\frac{1}{2}\frac{d^2}{d (\ln g)^2}
+ (1+\tilde{X}^M(g))\frac{d}{d \ln g}~]
+ \sum_\l m_\l^2 \tilde{X}^\l(g)\frac{d}{d \ln g} \}~\gamma_i(g)~,
\label{bm21}
\ee
which vanishes if $\gamma_i(g)=0$.
Therefore eq. (\ref{sum1})
with (\ref{F}) is the desired sum rule for the finite theories.
Since in two-loop order
$(\ln Y^{ijk})'=1$, $(\ln Y^{ijk})''=0$
and $X$ is given by eq. (\ref{x2}), we 
reproduce our previous result \cite{kkmz1}
\be
m_i^2+m_j^2+m_k^2 &=&|M|^2+\tilde{X}^{(2)}~,
\ee
where $\tilde{X}^{(2)}$ \cite{jack1,jack5} is given in (\ref{x2}).
The general case is more involved.
Following ref. \cite{jack4}  we require that the sum rule
(\ref{sum1}) with $F^M$ and $F^\l$ given in (\ref{F}) is RG
invariant in the general case, too. 
That is,   the reduction equation 
of the form \cite{kmz2}
\be
{\cal D}[~m^2_i+m^2_j+m^2_k -
|M|^2 F^M_{ijk}(g)
-\sum_\l m^2_\l F^{\l}_{ijk}~]& =& 0
\ee
has to be  satisfied,
where 
\be
{\cal D} &\equiv& \beta_g \frac{\partial}{\partial g}
+\beta_M \frac{\partial}{\partial M}
+\beta_M^* \frac{\partial}{\partial M^*}
+\sum_\l\beta_{m^2_\l} \frac{\partial}{\partial m^2_\l}~.
\ee
The equation above implies that
\be
& &\beta_{m^2_i}+\beta_{m^2_j}+\beta_{m^2_k}\nn\\
&=&
 \{|M|^2[~
\frac{1}{2}\frac{d^2}{d (\ln g)^2}
+ (1+\tilde{X}^M(g))\frac{d}{d \ln g}~]
+ \sum_\l m_\l^2 \tilde{X}^\l(g)\frac{d}{d \ln g} \}~
[\gamma_i(g)+\gamma_j(g)+\gamma_k(g)~]\nn\\
&=&
|M|^2\{~2(\beta_g /g)'~[(1+\tilde{X}^M)(\ln Y^{ijk})'
+\frac{1}{2}(\ln Y^{ijk})'']\nn\\
& &+(\beta_g /g)~[(\tilde{X}^M)'(\ln Y^{ijk})'
+(1+\tilde{X}^M)(\ln Y^{ijk})''
+\frac{1}{2}(\ln Y^{ijk})''')\nn\\
& &+\sum_\l \tilde{X}^\l (\ln Y^{ijk})'
[~\frac{1}{2}
(\gamma_{\l})''+
(1+\tilde{X}^M)(\gamma_{\l})']\nn\\
& &+\sum_\l m^2_\l \{~(\beta_g /g) 
[(\tilde{X}^\l)' (\ln Y^{ijk})'+\tilde{X}^\l (\ln Y^{ijk})'']
+ \tilde{X}^\l (\ln Y^{ijk})' \sum_m (\gamma_m)' \tilde{X}^m ~\}~,
\label{3beta}
\ee
where use has been made 
of eqs. (\ref{betaM}), (\ref{betam2}), 
(\ref{lny}), (\ref{bm21}) and
\be
{\cal O} &=&\frac{1}{2}M \frac{d}{d \ln g}~.
\ee
The eq. (\ref{3beta}) is satisfied if
\be
[(\beta_g /g)\tilde{X}^M]'+\sum_\l \tilde{X}^\l[
~\frac{1}{2}(\gamma_\l)''+(1+\tilde{X}^M)(\gamma_\l)']
&=&\frac{1}{2}(\beta_g /g)''-(\beta_g /g)'~,
\label{cond1}\\
\tilde{X}^i (\beta_g /g)'-(\tilde{X}^i )' (\beta_g /g)&=&
\tilde{X}^i \sum_\l \tilde{X}^\l (\gamma_\l)'~
\label{cond2}
\ee
are satisfied. It seems a highly non trivial task to solve these nonlinear
ordinary differential equations. 
On the other hand, there is another constraint coming from
the result of \cite{jack4},
given in eq. (\ref{xg}), for which it is assumed
 that
$m^2_i$ are also reduced in favor of
$g$ and $M$: It reads
\be
\sum_\l \tilde{X}^\l (\gamma_\l)'
&=&
-2(1+ \tilde{X}^M)(\beta_g /g)+(\beta_g /g)'~.
\label{cond3}
\ee
For a given 
$\beta_g$, it may be in principle possible to
solve eqs. (\ref{cond1}), (\ref{cond2}) together 
with the constraint (\ref{cond3})
to find $\tilde{X}^M(g)$ and $\tilde{X}^\l(g)$.
We find that  this
set of non-linear differential equations 
can be  solved for
the $\beta$ function of Novikov {\em et al.} 
\cite{novikov1} which is given
by \be
\beta_g^{\rm NSVZ} &=& 
\frac{g^3}{16\pi^2} 
\left[ \frac{\sum_\l T(R_\l)(1-\gamma_\l /2)
-3 C(G)}{ 1-g^2C(G)/8\pi^2}\right]~, 
\label{bnsvz}
\ee 
because $ \beta_g^{\rm NSVZ}$ has a 
certain singularity at 
\be
g^2 &=& \frac{8\pi^2}{C(G)}~.
\ee
We assume that $\tilde{X}^M$ and $\tilde{X}^\l$
 have a singularity of the form
\be
\tilde{X}^M &\sim& (C(G)-8\pi^2/g^2)^{-a}~,\nn\\
\tilde{X}^\l &\sim& (C(G)-8\pi^2/g^2)^{-a_\l}~,
\ee
and that $\gamma_\l(g)$ has no singularity
at $g^2 =8\pi^2/C(G)$. To find 
$a$ and $a_\l$
we  derive from eqs. (\ref{cond2}) and (\ref{cond3}) 
\be
(\ln \tilde{X}^\l)' &=& \tilde{X}^M+1 ~
\label{x+1}
\ee
which requires  that $a=1$. From eq. (\ref{cond3}) we find that
\be
1 \leq a_\l \leq 2~.
\ee
Further we find from eqs. (\ref{cond1}) and (\ref{cond3}) that
the leading singularity should be canceled without
the $\tilde{X}^\l$ terms in these equations, which fixes
$a_\l$ also to be one.
It is then straightforward to find the desired solution:
\be
\tilde{X}_{\rm NSVZ}^M&=&
-\frac{C(G)}{C(G)-8\pi^2/g^2}~,
\label{xm}\\
\tilde{X}_{\rm NSVZ}^\l &=&
\frac{T(R_\l)}{C(G)-8\pi^2/g^2}~,
\label{xl}
\ee
where we have used
\be
\sum_\l \gamma^{\rm NSVZ}_\l T(R_\l)
&=& (\beta^{\rm NSVZ}_g/g)
(C(G)-\frac{8\pi^2}{g^2})+
\frac{1}{2}[~\sum_\l T(R_\l)-3 C(G)~]~.
\label{gammaT}
\ee
Therefore, the sum rule (\ref{sum1}) in the NSVZ scheme
takes form
\be
m^2_i+m^2_j+m^2_k &=&
|M|^2 \{~
\frac{1}{1-g^2 C(G)/(8\pi^2)}\frac{d \ln Y^{ijk}}{d \ln g}
+\frac{1}{2}\frac{d^2 \ln Y^{ijk}}{d (\ln g)^2}~\}\nn\\
& &+\sum_\l
\frac{m^2_\l T(R_\l)}{C(G)-8\pi^2/g^2}
\frac{d \ln Y^{ijk}}{d \ln g}~.
\label{sum2}
\ee
This result should be compared with the superstring inspired result
for the finite case \cite{kkmz1} (i.e. 
$3C(G)=T(R)=\sum_\l T(R_\l)$)
\be
m^2_i+m^2_j+m^2_k &=&
|M|^2~
\frac{1}{1-g^2 C(G)/(8\pi^2)}+\sum_\l
\frac{m^2_\l T(R_\l)}{C(G)-8\pi^2/g^2}~,
\label{sum3}
\ee
which is valid in a certain class of orbifold models
 in which the massive string states are organized 
into $N = 4$ supermultiplets,
so that they do not contribute to 
the quantum modification of the kinetic
function \cite{kubo1}. 
So if $ (\ln Y^{ijk})'=1$,
the RG invariant expressions 
(\ref{h}) and (\ref{sum3}) exactly coincide
with the corresponding ones in the superstring models
in this particular case.

As a byproduct we obtain
the exact $\beta$ function for $m^2$
in the NSVZ scheme:
\be
\beta_{m^2_i}^{\rm NSVZ} &=&\left[~
|M|^2 \{~
\frac{1}{1-g^2 C(G)/(8\pi^2)}\frac{d }{d \ln g}
+\frac{1}{2}\frac{d^2 }{d (\ln g)^2}~\}\right.\nn\\
& &\left. +\sum_\l
\frac{m^2_\l T(R_\l)}{C(G)-8\pi^2/g^2}
\frac{d }{d \ln g}~\right]~\gamma_{i}^{\rm NSVZ}~,
\label{bm23}
\ee
where we have used eq. (\ref{bm21}),
(\ref{xm}) and (\ref{xl}).
Note that $\beta_{m^2_i}^{\rm NSVZ}$
assumes the form given in the r.h.s. of eq. (\ref{bm23})
only on the RG invariant surface defined
by $Y=Y(g)$ and eq. $(\ref{h})$ in the space
of parameters.
In theories without Yukawa couplings such as
supersymmetric QCD, the $\beta$ function above
is valid in the unconstrained space of parameters.

\section{Conclusions}

In the present paper we have shown to all orders in
perturbation theory the existence of the RGI
sum rule (\ref{sum1}) for
the soft scalar masses in the SSB sector of $N=1$ supersymmetric gauge
theories exhibiting gauge-Yukawa unification.
The all-loop sum rule (\ref{sum1}) with (\ref{F})
substitutes the universal soft scalar masses (which leads to phenomenological
problems), while the previously known relation among $h$'s, $Y$'s 
$M$ and $g$ still hold to all-loop \cite{kazakov1,jack4}.
 Particularly interesting is the
fact that the
finite unified theories, which could be made all-loop finite in the
supersymmetric sector \cite{lucchesi1,ermushev1,finite1}
 can be made completely 
{\em finite}, i.e. including the SSB sector,
in terms of the soft scalar-mass sum rule (\ref{sum1}),
generalizing the recent result of Kazakov \cite{kazakov1}
and relaxing his finiteness conditions.

This very appealing theoretical result complements nicely the successful
earlier prediction of the top quark mass 
\cite{finite1,kmz1,kmoz1} and the recent prediction
of the Higgs masses and the s-spectrum 
\cite{kkmz1}.

In the NSVZ scheme, the sum rule can be written in a more explicit
form (see (\ref{sum2})), exhibiting a definite singularity
at $g^2=8\pi^2/C(G)$.
The same singular behavior in the exact sum rule (\ref{sum3})
in a certain class of 
superstring models has been observed \cite{kkmz1}.
This result seems to be  suggesting a  hint for
 a possible connection among the
two kinds of theories.

\end{document}